# Room temperature synthesis of gallium oxide film with a fluidic exfoliation method

Fengyu Xu[1], Jianyu Wang[1*], Li Wang[1]

[1]*Department of Physics, Nanchang University, Nanchang, 330031, China*

*jywang@ncu.edu.cn

**Abstract**

Two-dimensional metal oxides play an important role in electronics and optoelectronics, and it is still a challenge to obtain thin oxides film. Here, a fluidic exfoliation method is applied to synthesis the metal oxides film by using galinstan as the reactant, and $Ga_2O_3$ film with ~1 cm size is obtained. Optical microscope and scanning electron microscope images show that the $Ga_2O_3$ film is exfoliated from the galinstan without any droplets left. Energy Dispersive X-Ray measurements confirm the existence of the $Ga_2O_3$ film. Transmission electron microscope and selected area electron diffraction patterns indicate the oxidation process do not have a prior direction. The alloy liquid based fluidic exfoliation method in room temperature provide a promising route for the synthesis of two-dimensional mental oxides, which shows significant applications in electronic and photoelectronic devices.

## 1. Introduction

Two-dimensional metal oxides are promising materials in electronic [1-3] and photoelectronic [4, 5] devices such as sensors and detectors [6-8]. $Ga_2O_3$ shows its advantages due to a large bandgap [9, 10]. For example, based on good ultraviolet transmittance, stability and wide bandgap (4.9 eV), $Ga_2O_3$ is potentially used for deep ultraviolet transparent conductor oxides (TCO) [6, 11, 12]. Gallium oxide is a natural material for solar-blind photodetector due to its cut-off wavelength (250~280 nm) [7, 13, 14]. $Ga_2O_3$ can also serve as a reactive oxide layer, sensitive to a wide variety of gases, is a promising choice for gas sensing applications at high temperatures and in harsh environments [6, 15-18]. In general, edge-defined film fed [19], floating-zone [20], and Czochralski [21, 22] methods are used to synthesize $Ga_2O_3$ thin film. Recently, Ali Zavabeti et al [23] reported a liquid mental reaction method and successfully obtained atomically thin layer, which can be used in the synthesis of various oxide nanomaterials with low dimensionality.

In this work, we applied a fluidic exfoliation method with galinstan alloy to synthesize the $Ga_2O_3$ film. Optical microscope and scanning electron microscope (SEM) images show the morphologies of $Ga_2O_3$ thin film without any droplets. Energy Dispersive X-Ray (EDX) measurement indicate the existence of $Ga_2O_3$. Transmission electron microscope (TEM) and selected area electron diffraction (SAED) results reveal the crystalline quality of $Ga_2O_3$ film. The fluidic exfoliation method with alloy liquid is useful for room temperature synthesis of metal oxide film, which is significantly important for electronic device applications.

## 2. Experimental

The synthesis process was carried out in a glove box with an oxygen concentration of 0.1 ppm. The galinstan liquid (Tin signet, 68.5% gallium, 21.5% indium, and 10% tin) is used as the reactant. A polyethylene funnel (the schematic shown in **Fig. 1(a)**) is applied to help with the formation of $Ga_2O_3$ film, and **Fig. 1(b)** is the optiacal image of the polyethylene funnel. During the synthesis process, a drop of fresh galinstan was placed on the funnel. Only the top surface of galinstan was exposed to the oxygen and oxidized slowly. With the galinstan flowing down through the funnel, oxidized $Ga_2O_3$ film was left on the funnel surface, as shown in **Fig. 1(c)**. **Fig. 1(d)** is the optical image of the left $Ga_2O_3$ film.

The structure and morphology of $Ga_2O_3$ film is characterized by optical microscope (Olympus BX23M), and SEM (Phenom Pro). TEM (JEM-2100) and SAED measurement is applied to investigate the crystalline quality. EDX is used to study the element composition of the $Ga_2O_3$ film.

## 3. Results and discussions

The oxidation process firstly carried out on the galinstan surface since it is exposed to the oxygen. As $Ga_2O_3$ has a lower Gibbs energy compared to the other elements in galinstan, the formation of $Ga_2O_3$ is prior to other oxides [24]. The formed $Ga_2O_3$ film will prevent the oxygen from entering the galinstan, which will reduce the oxidization speed [25, 26]. Therefore, the $Ga_2O_3$ film will left with a thin layer. Under the optical microscope, the film in the funnel is easily to be observed, as shown in **Fig. 2**. In **Fig. 2(a)**, the boundaries of the film are not very obvious, implying the oxide film is very thin with a few layers. In other area of **Fig. 2(b)**, the image shows much thicker films with hundreds of layers, indicating that the film thickness is not uniform. The van der Waals exfoliation method [27, 28] is also applied as a comparison, and the optical image is shown in **Fig. 2(c)**. Large galinstan droplets are left on the $Ga_2O_3$ film, which make it difficult to the transfer and further applications [29, 30]. By using the fluidic exfoliation method, pure $Ga_2O_3$ film can be obtained without any galinstan droplets.

For a detailed investigation on the morphologies, silicon substrate is used to transfer the $Ga_2O_3$ film, and the SEM image is shown in **Fig. 2(d)**. In the insert of **Fig. 2(d)**, clear film boundaries are observed, confirming the existence of the film, the black spots on the film are attributed to galinstan, oxide multilayer or other impurities during the transfer process.

The synthesized $Ga_2O_3$ film is transferred to Cu mesh and investigated with TEM, as shown in **Fig. 3**. In **Fig. 3(a)**, the EDX elemental analysis, Ga and O elements are observed, confirming the existence of $Ga_2O_3$. The high intensity of Cu elements is attributed to the Cu mesh. TEM image of the transferred $Ga_2O_3$ film is shown in **Fig. 3(b)**, in which the $Ga_2O_3$ film is clearly observed. The dark spots are also discovered in the TEM image, corresponding to the impurities during the transfer process. The SAED patterns in **Fig. 3(c)** show the $Ga_2O_3$ film is polycrystalline, indicating the oxidation of Ga is randomly orientated and the synthesized $Ga_2O_3$ film do not have a priority direction.

It is noted that the oxidation time is highly depended on the temperature, which influences the thickness and properties of the $Ga_2O_3$ film. Therefore, the oxidation time with varied temperature is investigated. The temperature was controlled by a hot plate, and exfoliation was performed every 1 minute to get the exact oxidation time. The results are shown in **Fig. 4**, and the inserts are the optical images of the exfoliated $Ga_2O_3$ films. When the temperature is lower than 20°C, the oxidation time will cost more than 1h. With the temperature increase to 40°C, the oxidation time decreases to 5 minutes, which is difficult for exfoliation. Therefore, the appropriate oxidation time for the fluidic exfoliation method is between 20°C to 40°C.

## 4. Conclusions

In conclusion, we successfully synthesized $Ga_2O_3$ film through fluidic exfoliation method. Optical and SEM images show the pure $Ga_2O_3$ films without any galinstan droplets are obtained. EDX elemental analysis demonstrates the existence of $Ga_2O_3$ film. TEM image and SAED patterns indicate the oxidation of $Ga_2O_3$ does not have a prior orientation. The oxidation time of galinstan is investigated, which decreases with the increasing temperature. The fluidic exfoliation method provides potential routes for the synthesis of metal oxides, showing important applications in electronic and photoelectronic devices.

**Acknowledgement**

The authors acknowledge the financial support from National Natural Science Foundation of China under Grant No. 62005112.

**Declaration of Competing Interest**

None.

**Figure Captions**

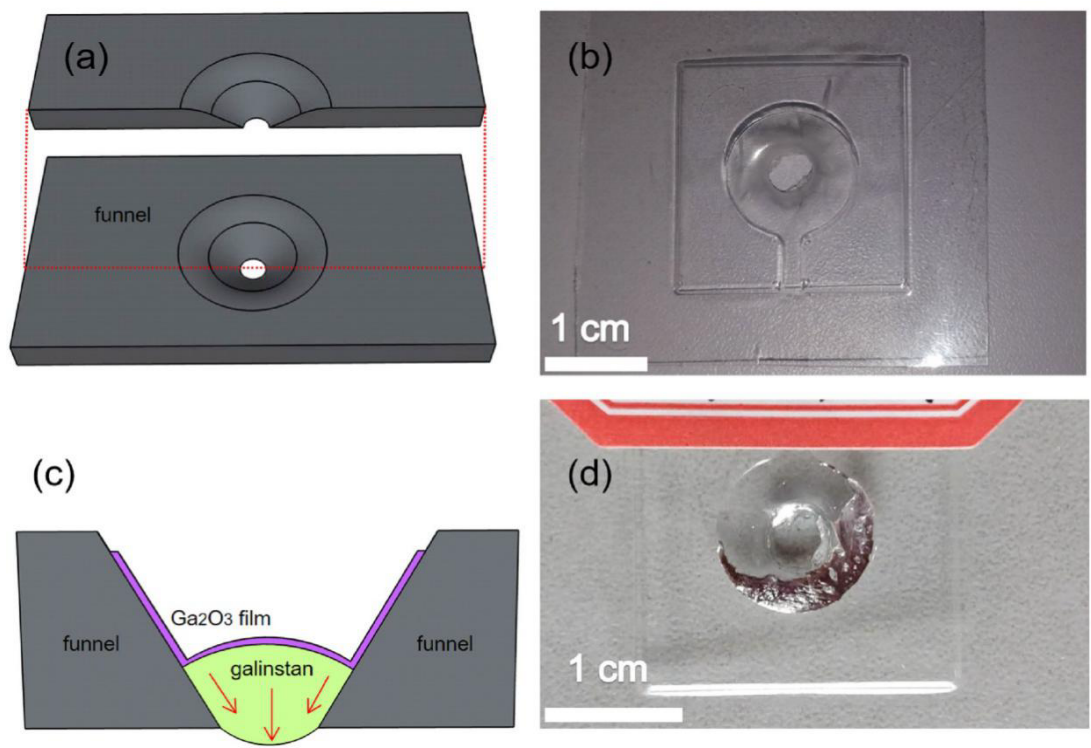

**FIG. 1.** (a) Schematic and (b) optical image of the polyethylene funnel. (c) Schematic of the $Ga_2O_3$ film synthesis process. (d) Optical image the synthesized $Ga_2O_3$ film.

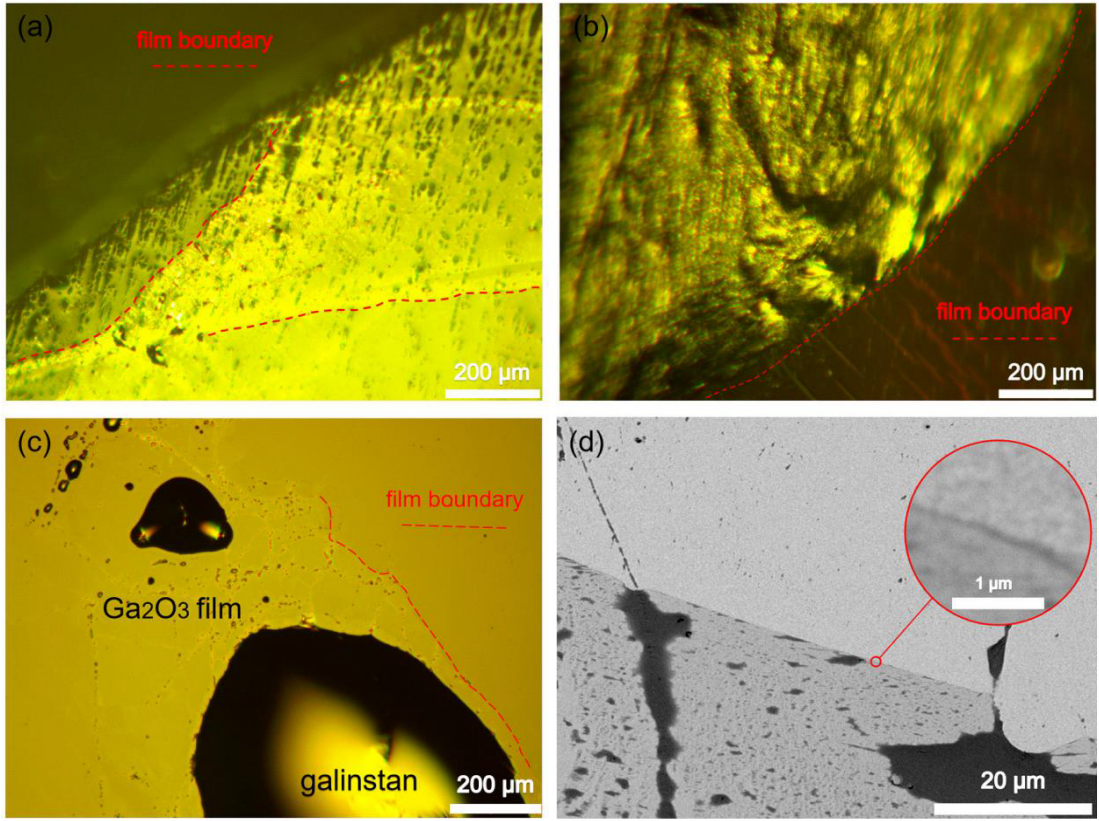

**FIG. 2.** (a), (b) Optical images of thin $Ga_2O_3$ film from fluidic exfoliation method with different thickness. (c) Optical image of $Ga_2O_3$ film with traditional van der Waals exfoliation method. (d) SEM image of the $Ga_2O_3$ film transferred to Si substrate.

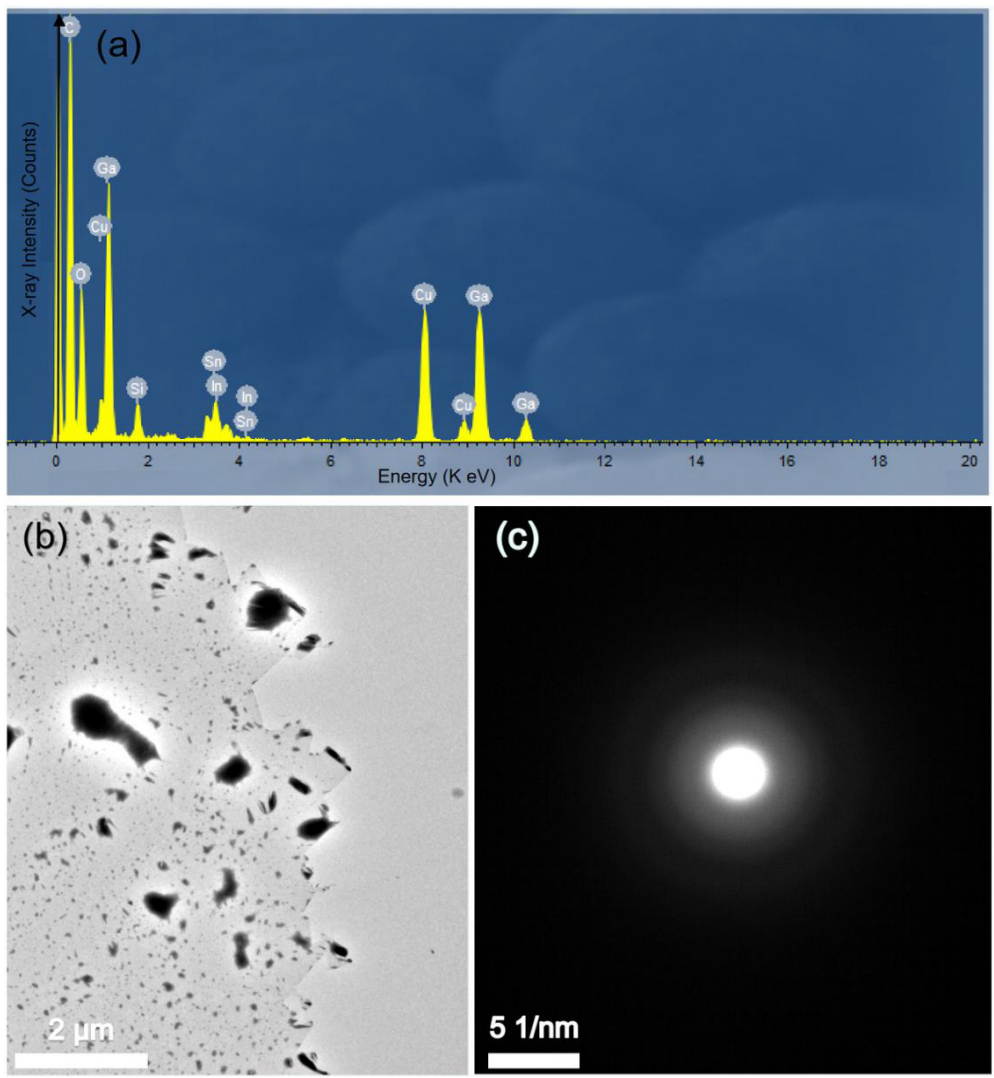

**FIG. 3.** (a) EDX spectra of the $Ga_2O_3$ film transferred to Cu mesh. (b) TEM image of the $Ga_2O_3$ film and (c) SAED patterns.

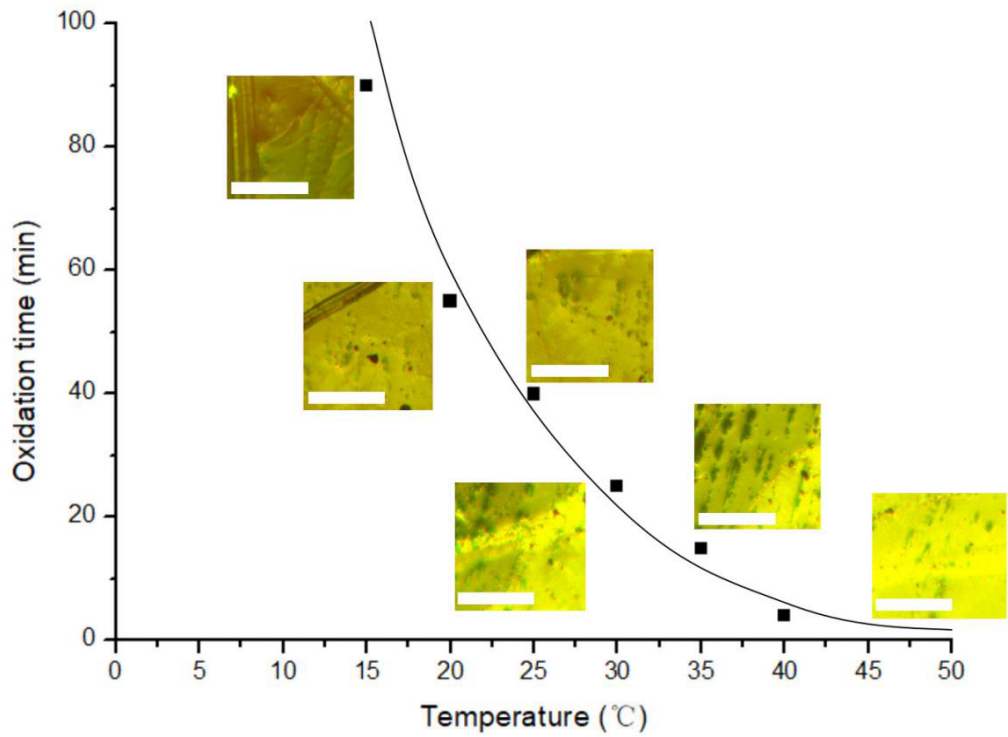

**Fig. 4.** Oxidation time diagram of $Ga_2O_3$ with varied temperatures, the inserts are the optical images of $Ga_2O_3$ film corresponding to the marked temperatures, all the scale bars in the inserts are 100 μm.